\documentclass[prd,preprintnumbers,superscriptaddress,tightenlines,nofootinbib, eqsecnum]{revtex4-2}

\usepackage{amsmath}
\usepackage{amsfonts}
\usepackage{amssymb}
\usepackage{bm}
\usepackage[colorlinks]{hyperref}
\usepackage{mathrsfs}
\usepackage{graphicx}
\usepackage{empheq}
\usepackage{ulem}
\usepackage{tensor}
\usepackage[usenames]{color}
\definecolor{darkgreen}{rgb}{0,0.5,0}
\hypersetup{urlcolor=darkgreen}
\usepackage[capitalize]{cleveref}

\allowdisplaybreaks

\DeclareSymbolFontAlphabet{\mathrsfs}{rsfs}
\DeclareMathAlphabet{\mathcal}{OMS}{cmsy}{m}{n}

\newcommand{\calO}{\mathcal{O}}
\newcommand{\dd}{\mathrm{d}}
\newcommand{\dM}{\mathrm{M}}
\newcommand{\dS}{\mathrm{S}}

\newcommand{\Ke}{K_\varepsilon(\tau)}
\newcommand{\qb}{\bar{q}}
\newcommand{\go}{\mathfrak{g}}

\begin{document}
	
\title{Conservative (failed)-tail effects at the fifth post-Newtonian order}

\author{Quentin \textsc{Henry}}\email{quentin.henry@aei.mpg.de}
\affiliation{Max Planck Institute for Gravitational Physics\\ (Albert Einstein Institute), D-14476 Potsdam, Germany}

\author{Fran\c{c}ois \textsc{Larrouturou}}\email{francois.larrouturou@desy.de}
\affiliation{Deutsches Elektronen-Synchrotron DESY, Notkestr. 85, 22607 Hamburg, Germany}

\date{\today}

\preprint{DESY-23-100}

\begin{abstract}

This work deals with the tail and ``failed'' tail sectors of the conservative dynamics for compact binary systems at the 5PN order.
We employ the Fokker Lagrangian method with dimensional regularization, and our results for the tail sector are perfectly consistent with the previous EFT computations.
As for the ``failed'' tail sector, we have good hopes that this new computation will help solving the current discrepancy in the literature.

\end{abstract}

\maketitle

\section{Introduction}
\label{sec:intro}

The post-Newtonian (PN) approximation scheme is a very efficient framework that relies on  weak-field and slow-velocity approximations to perturbatively solve Einstein's equation.
It has been notably implemented through a large class of methods to resolve the dynamics of bound compact binaries system, see \emph{e.g.}~\cite{BlanchetLR,Buonanno:2014aza,Porto:2016pyg} for reviews.
In the conservative sector,\footnote{This paper focuses on the conservative sector, \emph{i.e.} the study of the (conserved) dynamics of the system.
Nevertheless, those PN frameworks are also in use to solve the dissipative sector, \emph{i.e.} to derive the waveform.
Notably, using matched post-Newtonian and multipolar-post-Minkowskian methods~\cite{Blanchet:1998in}, the gravitational flux at 4PN and phase at 4.5PN were recently obtained~\cite{Blanchet:2023sbv,Blanchet:2023bwj}. 
Note also that the 2PN sector of the gravitational flux has been confirmed by EFT means~\cite{Leibovich:2019cxo}.} the current accuracy is the fourth PN order (\emph{i.e.} the $(v/c)^8$ correction to the Newtonian energy and angular momentum), that was obtained by means of the canonical Hamiltonian formalism~\cite{Jaranowski:2013lca,Jaranowski:2015lha,Damour:2014jta}, the Fokker method~\cite{Bernard:2015njp,Bernard:2016wrg,Marchand:2017pir,Bernard:2017bvn,Bernard:2017ktp}, and effective field theory (EFT) approach~\cite{Goldberger:2004jt,Foffa:2011np,Foffa:2012rn,Galley:2015kus,Foffa:2016rgu,Foffa:2019rdf,Foffa:2019yfl,Blumlein:2020pog}.
Starting at this 4PN precision, the conservative dynamics can be split between an ``instantaneous'' sector and a ``hereditary'' one.
The latter takes into account the back-reaction of emitted radiation onto the dynamics of the binary, which induces non-local in time effects (thus the name).
The computation of the instantaneous dynamics has been completed at 5PN by a large variety of methods~\cite{Foffa:2019hrb,Bini:2019nra,Bini:2020wpo,Blumlein:2020pyo}, and pushed up to the 6PN precision~\cite{Bini:2020nsb} (see also~\cite{Bini:2020rzn}).
As for the hereditary sector, due to the very subtle nature of the computations, only partial results exist.
For instance, the tail sector (due to the scattering of waves onto the static curvature induced by the ADM mass) has been computed by means of EFT~\cite{Almeida:2021xwn}.
As for the ``failed'' tail\footnote{We borrow this nomenclature to~\cite{Foffa:2019eeb}, where it has been dubbed ``failed'' as, although it comes as an hereditary effect, this interaction fails to induce a non-local-in-time sector.} (due to the scattering of waves onto the static curvature induced by the ADM angular momentum), it was also derived within the EFT framework. However, there is a discrepancy between previous results~\cite{Foffa:2019eeb,Blumlein:2021txe} and the recent work of~\cite{Gabriel}.
Note also that the logarithmic dependencies in the binding energy (due to this hereditary sector) are known up to the 7PN order~\cite{Blanchet:2019rjs}. 

The aim of this work is to derive the (failed) tail effects by means of the Fokker method using dimensional regularization.
We thus work with $d=3 + \varepsilon$ space-like dimensions and the $d$-dimensional gravitational strength, $G$, is linked to the usual Newton constant $G_N$ by a new length scale $\ell_0$ as $G = \ell_0^\varepsilon G_N$ (this regularization constant is directly related to the scale $\mu$ used in EFT framework~\cite{Almeida:2021xwn,Gabriel} as $\ell_0 = \mu^{-1}$).
The tail interactions entering at 5PN involve the constant ADM mass $\dM$, the mass octupole $\dM_{ijk}$ and the current quadrupole $\dS_{i\vert jk}$ (as we work in $d$ dimensions, we use the notations and conventions of~\cite{Henry:2021cek} for current moments).
As for the failed tail, it describes the interaction between the constant angular mometum $\dS_{i\vert j}$ and the mass quadrupole $\dM_{ij}$.
Note that the $d$-dimensional current dipole, $\dS_{i\vert j}$, simply reduces in three dimensions to $\lim_{d \to 3}\dS_{i\vert j} = \varepsilon_{ijk}L^k$, where $L^i$ is the usual angular momentum.
Our result reads
\begin{equation}\label{eq:intro_res}
\begin{aligned}
\mathcal{S}_\text{tail} = 
& - \frac{G^2\,\dM}{c^{10}} \int\!\!\dd t\int_0^\infty\!\!\dd\tau\bigg\lbrace 
\frac{1}{189} \left(\Ke  - \frac{82}{35}\right) \dM_{ijk}^{(4)}(t)\,\dM_{ijk}^{(5)}(t-\tau)
+\frac{16}{45} \left( \Ke  - \frac{49}{20}\right) \dS_{k\vert ji}^{(3)}(t)\,\dS_{k\vert ji}^{(4)}(t-\tau)\bigg\rbrace \\
&
+ \frac{G^2}{30\,c^{10}}\int\!\!\dd t\,\dS_{i\vert j}\,\dM_{ik}^{(3)}(t)\,\dM_{jk}^{(4)}(t)\,,
\end{aligned}
\end{equation}
where parenthetical superscripts denote time derivations and we have dressed the pole as
\begin{equation}\label{eq:intro_Ke}
\Ke \equiv \frac{1}{\varepsilon} -2 \ln\left(\frac{c\sqrt{\qb} \,\tau}{2\,\ell_0}\right)\,,
\qquad\text{with}\qquad
\qb \equiv 4\pi \,e^{\gamma_E}\,,
\end{equation}
where $\gamma_E$ is the Euler constant.
The first line of~\eqref{eq:intro_res}, corresponding to the tail interactions, is in perfect agreement with Eqs. (5) and (9) of~\cite{Almeida:2021xwn}.
As for the failed tail (the second line), we fully agree with the recent result of~\cite{Gabriel}, obtained by an independent method.

The plan of this paper is as follows.
Sec.~\ref{sec:method} describes the method employed to derive the (failed) tail effects, namely the Fokker method with dimensional regularization.
This method is then applied to each interaction separately in Sec.~\ref{sec:results}.
Finally, Sec.~\ref{sec:concl} concludes our work.

\section{General method}\label{sec:method}

In order to perform the computation of the conservative (failed) tail sector at 5PN, we follow the method that was used for the lowest-order tail $\dM \times \dM_{ij} \times \dM_{ij}$, entering at 4PN~\cite{Bernard:2017bvn,Marchand:2017pir}.
The following section briefly recalls and discuss its main steps.\footnote{The conventions employed throughout this work are as follows: we work with a mostly plus signature; greek letters denote spacetime indices and latin ones, purely spatial indices; bold font denotes $d$-dimensional vectors, \emph{e.g.} $\boldsymbol{y}_A = y_A^i$; we use the multi-index notations of~\cite{Henry:2021cek} (coming from Young tableaux), \emph{i.e.} $\dM_L = \dM_{i_1i_2\ldots i_\ell}$ and $\dS_{i\vert L} = \dS_{i\vert i_\ell\ldots i_2i_1}$; hats and angular brackets denote a symmetric and trace-free operator, $\hat{x}_L = x_{\langle L\rangle} = \text{STF}[x_L]$; the d'Alembertian operator is defined with respect to the flat Minkowski metric $\Box \equiv \eta^{\mu\nu}\partial_{\mu\nu} = \Delta - c^{-2}\partial_t^2$; (anti-)symmetrizations are weighted, \emph{e.g.} $A_{(ij)} = (A_{ij} + A_{ji})/2$; the Lagrangian and Lagrangian density are denoted as $\mathcal{S} = \int\!\dd t\,\mathcal{L} =  \int\!\dd t\dd^d x\, L$, and we will refer to ``Lagrangian'' for ``Lagrangian density'' henceforth; finally, and as usual, we dubb ``$n$PN'' a quantity of order $\calO(c^{-2n})$.}

\subsection{Tail effects in the action}\label{sec:method_action}

The starting point of the method is naturally an action composed of two sectors: the gravitational kinetic term and the matter description.
For the first one, we work with the usual Landau-Lifschitz Lagrangian, together with a gauge-fixing term (see \emph{e.g.} \cite{Bernard:2015njp})
\begin{equation}\label{eq:method_action_LL}
\mathcal{S}_\text{g}
= \frac{c^4}{16\pi G} \int\!\!\dd t\dd^dx\,\sqrt{-g}\, \bigg[g^{\mu\nu}\left(\Gamma_{\mu\rho}^\lambda\Gamma_{\nu\lambda}^\rho-\Gamma_{\mu\nu}^\lambda\Gamma_{\rho\lambda}^\rho\right)- \frac{1}{2}\,g_{\mu\nu}\Gamma^\mu\Gamma^\nu\bigg]\,,
\end{equation}
where $\Gamma^\mu_{\nu\rho}$ are the Christoffel symbols and the last term enforces the gauge $\Gamma^\mu \equiv g^{\alpha\beta}\Gamma_{\alpha\beta}^\mu = 0$.
In terms of the so-called ``gothic metric'' $\go^{\mu\nu} \equiv \sqrt{-g}\,g^{\mu\nu}$, this action becomes
\begin{equation}\label{eq:method_action_grav}
\mathcal{S}_\text{g} = \frac{c^4}{32\pi G} \int\!\!\dd t\dd^dx\, \left[
\go_{\alpha\beta}\left(\partial_\mu \go^{\alpha\nu}\,\partial_\nu \go^{\beta\mu}-\partial_\mu\go^{\alpha\mu}\,\partial_\nu \go^{\beta\nu}\right)
- \frac{1}{2} \go^{\alpha\beta}\go_{\mu\nu}\go_{\sigma\tau}\left(
\partial_\alpha \go^{\mu\sigma}\,\partial_\beta \go^{\nu\tau}
- \frac{1}{d-1}\,\partial_\alpha \go^{\mu\nu}\,\partial_\beta \go^{\sigma\tau}\right)\right]\,.
\end{equation}
As for the matter sector, we consider structureless, non-spinning point-particles, thus described by the action
\begin{equation}\label{eq:method_action_mat}
\mathcal{S}_\text{pp} = -c\,\sum_A\,m_A \int\!\!\dd\tau_A
= -c^2\sum_A\,m_A\int\!\!\dd t\,\dd^dx\,\frac{\delta_A}{u_A^0}\,,
\end{equation}
where $m_A$ is the mass of the particle $A$, $\tau_A$ its proper time, $u_A^0 \equiv [-(g_{\mu\nu})_A\,v_A^\mu v_A^\nu/c^2]^{-1/2}$ is the associated Lorentz factor, $v_A^\mu = (c,v_A^i)$ (with $v_A^i$ the usual velocity), and the $d$-dimensional Dirac distribution $\delta_A \equiv \delta[\boldsymbol{x} - \boldsymbol{y}_A(t)]$ locates the Lagrangian on the world-line of the particles.
As we are interested by the dynamics of binary systems, we will run $A$ only over two values.\\

From the gothic metric, we define the exact perturbation
\begin{equation}
h^{\mu\nu} \equiv \go^{\mu\nu} - \eta^{\mu\nu}\,,
\end{equation}
for which the gauge condition $\Gamma^\mu = 0$ translates into the usual harmonic gauge $\partial_\nu h^{\mu\nu} = 0$.
This perturbation obeys a wave equation source by the Laundau-Lifschitz pseudo-tensor $\tau^{\mu\nu}$
\begin{equation}\label{eq:method_h_waveEq}
\Box h^{\mu\nu} = \tau^{\mu\nu} = \frac{16\pi G}{c^4}\,\vert g\vert\, T^{\mu\nu}  + \Lambda^{\mu\nu}\,,
\end{equation}
where $T^{\mu\nu}$ is the stress-energy tensor of the matter distribution and $\Lambda^{\mu\nu}$ encrypts the non-linearities intrinsic to GR.
Its $d$-dimensional expression reads~\cite{BlanchetLR}
\begin{equation}\label{eq:method_Lambda}
\begin{aligned}
\Lambda^{\mu\nu} = & 
- h^{\alpha\beta}\partial_{\alpha\beta}h^{\mu\nu}
+ \partial_\alpha h^{\mu\beta}\partial_\beta h^{\nu\alpha}
+ \frac{1}{2}\,\go^{\mu\nu}\go_{\alpha\beta}\partial_\rho h^{\alpha\sigma}\partial_\sigma h^{\beta\rho}
-2\go^{\alpha(\mu}\go_{\rho\beta}\partial_\sigma h^{\nu)\beta}\partial_\alpha h^{\rho\sigma}\\
& + \go^{\alpha\beta}\go_{\rho\sigma}\partial_\alpha h^{\mu\rho}\partial_\beta h^{\nu\sigma}
+ \frac{1}{4}\left(2\go^{\mu\alpha}\go^{\nu\beta}-\go^{\mu\nu}\go^{\alpha\beta}\right)\left(\go_{\rho\sigma}\go_{\lambda\tau}-\frac{1}{d-1}\,\go_{\rho\lambda}\go_{\sigma\tau}\right)\partial_\alpha h^{\rho\lambda}\partial_\beta h^{\sigma\tau}\,.
\end{aligned}
\end{equation}

We are interested here in the near-zone behavior of the metric, \emph{i.e.} we aim at solving $\Box h^{\mu\nu}_\text{NZ} = \bar{\tau}^{\mu\nu}$, where $\bar{\tau}^{\mu\nu}$ is the PN expansion of $\tau^{\mu\nu}$.
The solution to such wave equation can be split in two sectors, as
\begin{equation}\label{eq:method_h_decomp}
h^{\mu\nu}_\text{NZ} = \bar{h}^{\mu\nu} + \mathcal{H}^{\mu\nu}\,.
\end{equation}
The first sector, $\bar{h}^{\mu\nu}$, is a particular solution of the wave equations, corresponding to the \emph{potential} modes of the EFT framework.
It is computed by applying the PN-expanded, regularized Green function on $\bar{\tau}^{\mu\nu}$, see \emph{e.g.} Eq.~(2.5) of~\cite{Marchand:2017pir}, and its expression is known up to 4PN~\cite{Marchand:2020fpt}. Due to PN expansions and regularization of the Green function, the metric $\bar{h}^{\mu\nu}$ is not computed using the correct prescription. Thus we have to add an homogeneous solution, $\mathcal{H}^{\mu\nu}$, in order to get the complete solution.
This solution is a consequence of the matching equation linking the near- and far-zone behaviors of the metric~\cite{Blanchet:1998in,BlanchetLR}, and its construction is the purpose of the next section.
As will be clear there, it corresponds to the conservative sector of the waves radiated by the source, and thus one can associate it to the \emph{radiative} modes of the EFT framework.\\

Following the spirit of the Fokker method, we inject the near-zone metric~\eqref{eq:method_h_decomp} into the conservative action~\eqref{eq:method_action_grav}--\eqref{eq:method_action_mat}, in order to obtain a resulting Lagrangian depending only on the matter variables (which accounts to integrating out the gravitational modes).
This yields an action mixing potential and radiative modes.
The sector free from any $\mathcal{H}^{\mu\nu}$ is the usual, instantaneous action, computed at 5PN by EFT means in \emph{e.g.}~\cite{Bini:2020wpo,Blumlein:2020pyo}, and we let its re-computation within the Fokker framework for future studies.
What interests us here is the linear-in-$\mathcal{H}^{\mu\nu}$ sector of the action, encompassing the leading order (failed) tail effects.\footnote{As the constant (ADM) masses and angular momentum do not radiate, the quadratic-in-$\mathcal{H}^{\mu\nu}$ sector of the action cannot contribute to tail effects at leading order. Note however that, at 5PN, this quadratic sector can contribute to the memory effect and to the 1PN corrections to the $\dM\times\dM_{ij}\times\dM_{ij}$ tail effect. The study of both those effects are left for future works.}
This linear sector can be interpreted as the backreaction of the scattered wave, $\mathcal{H}^{\mu\nu}$, onto the dynamics of the binary, thus describing indeed a tail effect.
This point of view corresponds to the closure of radiative Feynamm diagrams, performed in~\cite{Gabriel}.

As will be explicit hereafter, the different components of the radiative metric at a given PN order will follow $\mathcal{H}^{\mu\nu} = \calO(c^{-2n-2},c^{-2n-1},c^{-2n})$ with $\mathcal{H}^{kk} = \calO(c^{-2n-2})$ (in particular, $n=5$ for this work). The leading PN order of the linear-in-$\mathcal{H}^{\mu\nu}$ sector of the action reads
\begin{equation}\label{eq:method_action_LO}
\mathcal{S}^\text{tails}_\text{LO}
 = -\int\!\!\dd t\dd^dx \bigg\lbrace\frac{m_1\,c^2}{8} \bigg[
\mathcal{H}^{00ii}
- \frac{4\,v_1^i}{c}\,\mathcal{H}^{0i}
+ \frac{2\,v_1^{ij}}{c^2}\,\mathcal{H}^{ij}\bigg]\,\delta_1
+ \frac{(d-1)\,\mathcal{H}^{ij}}{64(d-2)\pi G}\,\partial_iV\partial_jV\bigg\rbrace + (1 \leftrightarrow2)\,,
\end{equation}
where we have shortened $\mathcal{H}^{00ii} \equiv \frac{2}{d-1}[(d-2)\mathcal{H}^{00}+\mathcal{H}^{ii}]$.
From the compact terms (proportional to the Dirac distribution), one will be able to reconstruct the Newtonian value of the moments.\footnote{For example, if $\mathcal{H}^{00ii} = \hat{x}^{ijk}F_{ijk}(t)$, where $F_{ijk}$ only depends on time, then $\int\!\dd^dx \,m_1 \mathcal{H}^{00ii}\delta_1 + (1\leftrightarrow 2) = O_{ijk} F_{ijk}$, where $O_{ijk} = m_1 \hat{y}_1^{ijk} + m_2 \hat{y}_2^{ijk}$ is the Newtonian mass octupole moment.}
As for the non-compact term (the last piece), it is treated by using the generalized Riesz formulae displayed in App. A of~\cite{Larrouturou:2021dma}.

\subsection{Computation of the radiative metric}\label{sec:method_hrad}

The radiative metric $\mathcal{H}^{\mu\nu}$ corresponds to an homogeneous solution of the wave equations~\eqref{eq:method_h_waveEq}, regular in the source (when $r \to 0$).
This means that it has the structure
\begin{equation}\label{eq:method_Hmunu_gen}
\mathcal{H}^{\mu\nu} = \sum_{j,\ell \geq 0}\,\Delta^{-j}\hat{x}_L\left(\frac{\dd}{c\,\dd t}\right)^{2j}\!\! f_L^{\mu\nu}(t)\,,
\qquad\text{with}\qquad
\Delta^{-j}\hat{x}_L = \frac{\Gamma\left(\frac{d}{2}+\ell\right)}{\Gamma\left(\frac{d}{2}+\ell+j\right)}\frac{r^{2j}\hat{x}_L}{2^{2j}\,j!}\,,
\end{equation}
where the functions $f_L^{\mu\nu}(t)$ are determined by the matching equation~\cite{Blanchet:1998in,BlanchetLR}, \emph{i.e.} by imposing that the near- and far-zone expansions of the metric agree in some overlapping region.
Therefore it is clear that $\mathcal{H}^{\mu\nu}$ encodes the fact that the dynamics of the system is sensitive to the gravitational waves radiated at spatial infinity.

This fact is even more evident when looking at the practical computation of $\mathcal{H}^{\mu\nu}$.
As derived in~\cite{Bernard:2017bvn}, the matching equation imposes that $\mathcal{H}^{\mu\nu}$ is nothing but an homogeneous solution of the far-zone expansion of the wave equations~\eqref{eq:method_h_waveEq}.
It is thus sourced by the expansion of $\Lambda^{\mu\nu}$~\eqref{eq:method_Lambda} at spatial infinity\footnote{We consider compact binaries, and so a compact-supported matter stress-energy tensor: at spatial infinity, $\tau^{\mu\nu}$ reduces to $\Lambda^{\mu\nu}$.} and can be computed by means of the $d$-dimensional usual multipolar-post-Minkowskian algorithm.

This algorithm starts with the $d$-dimensional generalization of Thorne's linearized metric~\cite{Thorne:1980ru}, namely~\cite{Henry:2021cek}
\begin{subequations}\label{eq:method_hlin}
	\begin{align}
& h^{00}_1=
- \frac{4}{c^2}\sum_{\ell \geqslant 0}\frac{(-)^\ell}{\ell !} \,\hat{\partial}_L\,\widetilde{\mathcal{M}}_L\,,\\
&
h^{0i}_1 =
\frac{4}{c^3}\sum_{\ell \geqslant 1}\frac{(-)^\ell}{\ell !}\,\left[\hat{\partial}_{L-1}\,\widetilde{\mathcal{M}}_{iL-1}^{(1)}+\frac{\ell}{\ell+1}\hat{\partial}_L\widetilde{\mathcal{S}}_{i\vert L}\right]\,,\\
&
h^{ij}_1 =
- \frac{4}{c^4}\sum_{\ell \geqslant 2}\frac{(-)^\ell}{\ell !}\,\left[
\hat{\partial}_{L-2}\,\widetilde{\mathcal{M}}_{ijL-2}^{(2)}+\frac{2\ell}{\ell+1}\hat{\partial}_{L-1}\widetilde{\mathcal{S}}^{(1)}_{(i\vert \underline{L-1} j)}+\frac{\ell-1}{\ell+1}\hat{\partial}_L\widetilde{\mathcal{K}}_{ij\vert L}\right]\,.
\end{align}
\end{subequations}
Underlined indices are excluded from symmetrization, and we have introduced the notation
\begin{equation}\label{eq:method_Mtilded_def}
\widetilde{\mathcal{M}}_L(r,t) \equiv \frac{\tilde{k}}{r^{d-2}}\int_1^{+\infty}\!\!\dd y\,\gamma_\frac{1-d}{2}(y)\,\dM_L\left(t- \frac{yr}{c}\right)\,,
\end{equation}
where
\begin{equation}\label{eq:method_gamma_def}
\tilde{k} \equiv \frac{\Gamma\left(\frac{d-2}{2}\right)}{\pi^\frac{d-2}{2}} = 1 - \frac{\varepsilon}{2}\,\ln\qb + \calO(\varepsilon^2) \qquad \text{and} \qquad \gamma_k(z) \equiv \frac{2\sqrt{\pi}}{\Gamma(1+k)\Gamma(-\frac{1}{2}-k)}\,\big(z^2-1\bigr)^k
\end{equation}
is such that $\lim_{d\to3}\widetilde{\mathcal{M}}_L(r,t) = \dM_L(t-r/c)/r$.
Note the presence of the additional set of moments $K_{ij\vert L}$, which are a pure artifact of working in $d \neq 3$ dimensions~\cite{Henry:2021cek}.

For our practical purpose, we will only consider interactions between a static moment (either the ADM mass $\dM$ or angular momentum $\dS_{i\vert j}$) and a propagating one.
Therefore, injecting in $\Lambda^{\mu\nu}$~\eqref{eq:method_Lambda} the sectors of the linear metric~\eqref{eq:method_hlin} that are of interest for us, the quadratic sources are of the form
\begin{equation}\label{eq:method_N_linF}
N(\mathbf{x},t) = \hat{n}_L\,\frac{\ell_0^{q\varepsilon}}{r^{p+q\varepsilon}}\int_1^{+\infty}\!\!\dd z\,\gamma_\frac{1-d}{2}(z)\,z^k\,F\left(t-\frac{zr}{c}\right)\,,
\end{equation}
where $F(t)$ represents a product of $\{\dM,\dS_{i\vert j}\}$ with (temporal derivatives of) one of the moments $\{\dM_{ij}, \dM_{ijk}, \dS_{i\vert jk}\}$ and $(k,\ell,p,q)$ take natural integer values.
Following the computation performed in~\cite{Bernard:2017bvn}, we then define the homogeneous solution $\mathcal{U}^{\mu\nu}$ corresponding to the source~\eqref{eq:method_N_linF} as
\begin{equation}\label{eq:method_Hmunu_expl}
\mathcal{U} = \frac{(-)^{p+\ell}}{d+2\ell-2} \underset{B=0}{\text{PF}} \,\frac{\Gamma\left(q\varepsilon-B\right)}{\Gamma\left(p+\ell-1+q\varepsilon-B\right)}\,C^{k,p,q}_\ell\,\sum_{j\in\mathbb{N}}\Delta^{-j}\hat{x}_L\int_0^\infty\!\!\dd\tau\frac{\tau^{B-q\varepsilon}}{r_0^B}\frac{F^{(2j+\ell+p-1)}(t-\tau)}{c^{2j+\ell+p+q\varepsilon-B}}\,,
\end{equation}
where the PF operator corresponds to the \emph{finite part} operation when $B\rightarrow 0$~\cite{Blanchet:2000nu}, and $C^{k,p,q}_\ell$ reads
\begin{equation}\label{eq:method_Ckpql_def}
C^{k,p,q}_\ell \equiv
\int_1^{+\infty}\!\!\dd y\,\gamma_{\frac{1-d}{2}-\ell}(y)\int_1^{+\infty}\!\!\dd z\,\gamma_{\frac{1-d}{2}}(z)\,z^k(y+z)^{\ell-2+p+q\varepsilon-B}\,.
\end{equation}
These coefficients are generalizations for $q \in \mathbb{Z}$ of the ones introduced in~\cite{Bernard:2017bvn}, and can be computed following the lines of the App. D of that work.\footnote{In the case of the memory interaction, the two moments under consideration are propagating, and thus sources are of the form $N \propto \int\dd y\,\gamma_\frac{1-d}{2}(y)\,y^k\,F(t-yr/c)\int\dd z\,\gamma_\frac{1-d}{2}(z)\,z^m\,G(t-zr/c)$. In such cases, we were not able to write the homogeneous solution in a form as simple as~\eqref{eq:method_Hmunu_expl}, notably because there are no factorization similar to~\eqref{eq:method_Ckpql_def}. \label{temp}}

If $\mathcal{U}^{\mu\nu}$ is of the form~\eqref{eq:method_Hmunu_gen}, namely an homogeneous solution regular in the source, it is not yet the homogeneous solution $\mathcal{H}^{\mu\nu}$ that we seek. At this stage, $\mathcal{U}^{\mu\nu}$ has no reason to be divergenceless and thus does not verify in general the harmonic condition. So to construct the correct solution, we add to $\mathcal{U}^{\mu\nu}$ a suited homogeneous solution, $\mathcal{V}^{\mu\nu}$, which cancels its divergence, following the standard procedure described \emph{e.g.} in~\cite{Blanchet:1998in,BlanchetLR}
\begin{align}
\mathcal{U}^{\mu\nu}\quad \longrightarrow \quad \mathcal{V}^{\mu\nu} = H\left(\partial_\mu \mathcal{U}^{\mu\nu}\right)\quad \longrightarrow \quad  \mathcal{H}^{\mu\nu}\equiv \mathcal{U}^{\mu\nu}+\mathcal{V}^{\mu\nu}\,.
\end{align}
In this method, $\mathcal{V}^{\mu\nu}$ is uniquely determined \emph{via} the harmonicity algorithm given by Eqs.~(47)--(48) in~\cite{BlanchetLR} and dubbed $H$ here.
Note that a similar removal of the divergence was employed in the EFT computation of~\cite{Gabriel}, and was in fact a crucial step to obtain the failed tail.

Once the divergenceless $\mathcal{H}^{\mu\nu}$ is known, one can inject it in the action~\eqref{eq:method_action_LO} and compute the integrals to obtain the desired effects.
In order to simplify the procedure, one can also implement a gauge transformation to ``push'' the $ij$ components of the metric to higher PN orders, and thus only have compact integrals to perform.
The metric transforms under the gauge transformation with vector $\xi^\mu$ as 
\begin{equation}
\mathcal{H}^{\mu\nu} \to \mathcal{H}'^{\mu\nu} = \mathcal{H}^{\mu\nu} + \partial^\mu\xi^\nu + \partial^\nu\xi^\mu - \partial_\rho \xi^\rho\,\eta^{\mu\nu} + \calO(\xi^2)\,.
\end{equation}
So, by choosing $\xi^\mu$ adequately, one can cancel the leading order of $\mathcal{H}^{ij}$.
In the next section, both raw and shifted metrics are displayed for each interaction, and we have naturally verified that they give the same result.

Note that we have also performed another consistency check on the results presented in the next section.
First, following the historical method developed in~\cite{Blanchet:1993ng}, we have implemented a purely Hadamard regularization procedure, yielding a three-dimensional metric, $\mathcal{H}^{\mu\nu}_\text{Had}$.
Then, using novel techniques elaborated in~\cite{Larrouturou:2021gqo} (and different from the one presented above), we have derived the contribution induced by the difference between the $d$-dimensional regularization scheme and the Hadamard one, \emph{i.e.} $\mathcal{D}\mathcal{H}^{\mu\nu} \equiv  \mathcal{H}^{\mu\nu}_{d-\text{dim}}- \mathcal{H}^{\mu\nu}_\text{Had}$.
By summing those two contributions, we recovered the metric computed directly in $d$ dimensions, \emph{i.e.} $\mathcal{H}^{\mu\nu}_\text{Had} + \mathcal{D}\mathcal{H}^{\mu\nu} = \mathcal{H}^{\mu\nu}$, which is a technically strong, although conceptually simple, check of our computations.

\section{Results at 5PN}\label{sec:results}

Let us implement the method described in the previous section (with extensive use of the \textit{xAct} library from the \textit{Mathematica} software~\cite{xtensor}) in the cases of the tails appearing at 5PN in the conservative action, composed of the $\dM \times \dM_{ijk} \times \dM_{ijk}$, $\dM \times \dS_{i\vert jk} \times \dS_{i\vert jk}$ and $\dS_{i\vert j} \times \dM_{ij} \times \dM_{ij}$ interactions.
We recall that pole is dressed as in Eq.~\eqref{eq:intro_Ke}.
By summing the separate results, we obtain our main result~\eqref{eq:intro_res}.

\subsection{Mass octupole tail}\label{sec:results_Oijk}

The divergenceless metric for the $\dM \times \dM_{ijk}$ interaction reads at the leading order
\begin{subequations}\label{eq:res_Hmunu_Oijk}
\begin{align}
\mathcal{H}^{00ii}_{\dM \times \dM_{ijk}} & = \frac{4\,G^2M\,\hat{x}^{ijk}}{315\,c^{12}} \int_0^\infty\!\!\dd\tau\left(\Ke - \frac{199}{70} \right)\dM_{ijk}^{(9)}(t-\tau)
+ \calO(c^{-14})\,,\\
\mathcal{H}^{0i}_{\dM \times \dM_{ijk}} & = -\frac{4\,G^2M\,\hat{x}^{jk}}{45\,c^{11}} \int_0^\infty\!\!\dd\tau\left(\Ke - \frac{1189}{420} \right)\dM_{ijk}^{(8)}(t-\tau)
+ \calO(c^{-13})\,,\\
\mathcal{H}^{ij}_{\dM \times \dM_{ijk}} & = \frac{4\,G^2M\,\hat{x}^{k}}{9\,c^{10}} \int_0^\infty\!\!\dd\tau\left(\Ke - \frac{113}{42} \right)\dM_{ijk}^{(7)}(t-\tau)
+ \calO(c^{-12})\,.
\end{align}
\end{subequations}
By applying the following shift
\begin{subequations}
\begin{align}
\xi^0_{\dM \times \dM_{ijk}} & = -\frac{G^2M\,\hat{x}^{ijk}}{189\,c^{11}} \int_0^\infty\!\!\dd\tau\left(\Ke - \frac{149}{70} \right)\dM_{ijk}^{(8)}(t-\tau)\,,\\
\xi^i_{\dM \times \dM_{ijk}} & =-\frac{G^2M\,\hat{x}^{ijk}}{9\,c^{10}} \int_0^\infty\!\!\dd\tau\left(\Ke - \frac{113}{42} \right)\dM_{ijk}^{(7)}(t-\tau)\,,
\end{align}
\end{subequations}
the metric becomes of order $\mathcal{H}'^{\ \mu\nu}_{\dM \times \dM_{ijk}} = \calO(c^{-12},c^{-13},c^{-12})$ and reads
\begin{equation}
\mathcal{H}'^{\ 00ii}_{\dM \times \dM_{ijk}} = \frac{8\,G^2M\,\hat{x}^{ijk}}{189\,c^{12}} \int_0^\infty\!\!\dd\tau\left(\Ke - \frac{82}{35} \right)\dM_{ijk}^{(9)}(t-\tau)
+ \calO(c^{-14})\,,
\end{equation}
which, injected in the action~\eqref{eq:method_action_LO}, yields (upon integrations by parts)
\begin{equation}\label{eq:res_Oijk}
\mathcal{S}_{\dM \times \dM_{ijk}} = -\frac{G^2M}{189\,c^{10}} \int\!\!\dd t \int_0^\infty\!\!\dd\tau\left(\Ke - \frac{82}{35} \right)\dM_{ijk}^{(4)}(t)\,\dM_{ijk}^{(5)}(t-\tau)
+ \calO(c^{-12})\,.
\end{equation}

\subsection{Current quadrupole tail}\label{sec:results_Sij}

The divergenceless metric for the $\dM \times \dS_{i\vert jk}$ interaction reads at the leading order
\begin{subequations}\label{eq:res_Hmunu_Sij}
\begin{align}
\mathcal{H}^{00ii}_{\dM \times\dS_{i\vert jk}} & = 
\calO(c^{-14})\,,\\
\mathcal{H}^{0i}_{\dM \times\dS_{i\vert jk}} & = \frac{8\,G^2M\,\hat{x}^{jk}}{45\,c^{11}} \int_0^\infty\!\!\dd\tau\left(\Ke - \frac{71}{30} \right)\dS_{i\vert jk}^{(7)}(t-\tau)
+ \calO(c^{-13})\,,\\
\mathcal{H}^{ij}_{\dM \times\dS_{i\vert jk}} & = -\frac{16\,G^2M\,\hat{x}^{k}}{9\,c^{10}} \int_0^\infty\!\!\dd\tau\left(\Ke - \frac{73}{30} \right)\dS_{(i\vert \underline{k}j)}^{(6)}(t-\tau)
+ \calO(c^{-12})\,.
\end{align}
\end{subequations}
By applying the following shift
\begin{equation}
\xi^0_{\dM \times\dS_{i\vert jk}} =0\,,
\qquad
\xi^i_{\dM \times\dS_{i\vert jk}} =\frac{8\,G^2M\,\hat{x}^{jk}}{9\,c^{10}} \int_0^\infty\!\!\dd\tau\left(\Ke - \frac{73}{30} \right)\dS_{i\vert jk}^{(6)}(t-\tau)\,,
\end{equation}
the metric becomes of order $\mathcal{H}'^{\ \mu\nu}_{\dM \times\dS_{i\vert jk}} = \calO(c^{-14},c^{-11},c^{-12})$, and reads
\begin{equation}
\mathcal{H}'^{\ 0i}_{\dM \times\dS_{i\vert jk}} = -\frac{32\,G^2M\,\hat{x}^{jk}}{45\,c^{11}} \int_0^\infty\!\!\dd\tau\left(\Ke - \frac{49}{20} \right)\dS_{i\vert jk}^{(7)}(t-\tau)
+ \calO(c^{-13})\,,
\end{equation}
which, injected in the action~\eqref{eq:method_action_LO}, yields (upon integrations by parts)
\begin{equation}\label{eq:res_Sij}
\mathcal{S}_{\dM \times\dS_{i\vert jk}} = -\frac{16\,G^2M}{45\,c^{10}} \int\!\!\dd t\int_0^\infty\!\!\dd\tau\left(\Ke - \frac{49}{20} \right)\dS_{i\vert jk}^{(3)}(t)\,\dS_{i\vert jk}^{(4)}(t-\tau)
+ \calO(c^{-12})\,.
\end{equation}

\subsection{Angular momentum failed tail}\label{sec:results_failed}

Finally, the divergenceless metric for the $\dS_{i\vert j} \times \dM_{ij}$ interaction reads at the leading order
\begin{subequations}\label{eq:res_Hmunu_failed}
\begin{align}
\mathcal{H}^{00ii}_{\dS_{i\vert j} \times \dM_{ij}} & = \frac{4\,G^2}{45\,c^{12}}\,\hat{x}^{jk}\,\dS_{i\vert k}^{}\, \dM_{ij}^{(7)}(t)
+ \calO(c^{-14})\,,\\
\mathcal{H}^{0i}_{\dS_{i\vert j} \times \dM_{ij}} & = \frac{4\,G^2}{9\,c^{11}}\,\hat{x}^{k}\,\dS_{j\vert k}^{}\, \dM_{ij}^{(6)}(t)
-\frac{G^2}{9\,c^{11}}\,\hat{x}^{k}\,\dS_{i\vert j}^{}\, \dM_{jk}^{(6)}(t)
+ \calO(c^{-13})\,,\\
\mathcal{H}^{ij}_{\dS_{i\vert j} \times \dM_{ij}} & = -\frac{22\,G^2}{15\,c^{10}}\,\dS_{k\vert (i}^{}\, \dM_{j)k}^{(5)}(t)
+ \calO(c^{-12})\,.
\end{align}
\end{subequations}
By applying the shift
\begin{subequations}
\begin{align}
\xi^0_{\dS_{i\vert j} \times \dM_{ij}} & = \frac{4\,G^2}{45\,c^{11}}\,\hat{x}^{jk}\,\dS_{i\vert k}^{}\, \dM_{ij}^{(6)}(t)\,,\\
\xi^i_{\dS_{i\vert j} \times \dM_{ij}} & = \frac{8\,G^2}{15\,c^{10}}\,\hat{x}^{k}\,\dS_{j\vert k}^{}\, \dM_{ij}^{(5)}(t)
-\frac{3\,G^2}{15\,c^{10}}\,\hat{x}^{k}\,\dS_{i\vert j}^{}\, \dM_{jk}^{(5)}(t)\,.
\end{align}
\end{subequations}
the metric becomes of order $\mathcal{H}'^{\ \mu\nu}_{\dS_{i\vert j} \times \dM_{ij}} = \calO(c^{-12},c^{-13},c^{-12})$, and reads
\begin{equation}
\mathcal{H}'^{\ 00ii}_{\dS_{i\vert j} \times \dM_{ij}} = - \frac{4\,G^2}{15\,c^{12}}\,\hat{x}^{jk}\,\dS_{i\vert k}^{}\, \dM_{ij}^{(7)}(t)\,
+ \calO(c^{-14})\,,
\end{equation}
which, injected in the action~\eqref{eq:method_action_LO}, yields (upon integrations by parts)
\begin{equation}\label{eq:res_Lij}
\mathcal{S}_{\dS_{i\vert j} \times \dM_{ij}} = \frac{G^2}{30\,c^{10}}\,\dS_{i\vert j}^{}\, \int\!\!\dd t\, \dM_{ik}^{(3)}(t)\,\dM_{jk}^{(4)}(t)
+ \calO(c^{-12})\,.
\end{equation}

\section{Summary and conclusion}\label{sec:concl}

In this work, we have derived the leading order tail and ``failed'' tail sectors appearing at the 5PN order in the conservative dynamics of compact binaries, by employing the Fokker Lagrangian framework.
Making use of dimensional regularization, we have computed the homogeneous solution of the near-zone metric, and have integrated it out in the action.
Our result, given in Eq.~\eqref{eq:intro_res}, is consistent with previous works performed within the EFT framework: the tail sector agrees with~\cite{Almeida:2021xwn}, and the failed tail one, with~\cite{Gabriel}.
With this new computation at hand, we hope that the current discrepancy in EFT results for the failed tail sector will be fully understood and resolved.

The last step towards completion of the 5PN conservative dynamics is the \emph{memory} effect, \emph{i.e.} the interaction of three mass quadrupoles.
In order to compute it within the Fokker Lagrangian framework, the method presented in this work has to be enhanced, as briefly discussed in the footnote~\ref{temp}.
This subtle computation is thus left for future work.

\acknowledgments

It is a pleasure to thank G. Luz Almeida, S. Foffa, A. M\"uller and R. Sturani for enlightening discussions at a late stage of this work, and Luc Blanchet at an early stage.
F.L. received funding from the European Research Council (ERC) under the European Union’s Horizon 2020 research and innovation program (grant agreement No 817791).

\bibliography{ListeRef_5PNtails.bib}

\begin{thebibliography}{42}%
\makeatletter
\providecommand \@ifxundefined [1]{%
 \@ifx{#1\undefined}
}%
\providecommand \@ifnum [1]{%
 \ifnum #1\expandafter \@firstoftwo
 \else \expandafter \@secondoftwo
 \fi
}%
\providecommand \@ifx [1]{%
 \ifx #1\expandafter \@firstoftwo
 \else \expandafter \@secondoftwo
 \fi
}%
\providecommand \natexlab [1]{#1}%
\providecommand \enquote  [1]{``#1''}%
\providecommand \bibnamefont  [1]{#1}%
\providecommand \bibfnamefont [1]{#1}%
\providecommand \citenamefont [1]{#1}%
\providecommand \href@noop [0]{\@secondoftwo}%
\providecommand \href [0]{\begingroup \@sanitize@url \@href}%
\providecommand \@href[1]{\@@startlink{#1}\@@href}%
\providecommand \@@href[1]{\endgroup#1\@@endlink}%
\providecommand \@sanitize@url [0]{\catcode `\\12\catcode `\$12\catcode
  `\&12\catcode `\#12\catcode `\^12\catcode `\_12\catcode `\%12\relax}%
\providecommand \@@startlink[1]{}%
\providecommand \@@endlink[0]{}%
\providecommand \url  [0]{\begingroup\@sanitize@url \@url }%
\providecommand \@url [1]{\endgroup\@href {#1}{\urlprefix }}%
\providecommand \urlprefix  [0]{URL }%
\providecommand \Eprint [0]{\href }%
\providecommand \doibase [0]{https://doi.org/}%
\providecommand \selectlanguage [0]{\@gobble}%
\providecommand \bibinfo  [0]{\@secondoftwo}%
\providecommand \bibfield  [0]{\@secondoftwo}%
\providecommand \translation [1]{[#1]}%
\providecommand \BibitemOpen [0]{}%
\providecommand \bibitemStop [0]{}%
\providecommand \bibitemNoStop [0]{.\EOS\space}%
\providecommand \EOS [0]{\spacefactor3000\relax}%
\providecommand \BibitemShut  [1]{\csname bibitem#1\endcsname}%
\let\auto@bib@innerbib\@empty
\bibitem [{\citenamefont {Blanchet}(2014)}]{BlanchetLR}%
  \BibitemOpen
  \bibfield  {author} {\bibinfo {author} {\bibfnamefont {L.}~\bibnamefont
  {Blanchet}},\ }\bibfield  {title} {\bibinfo {title} {{Gravitational Radiation
  from Post-Newtonian Sources and Inspiralling Compact Binaries}},\ }\href
  {https://doi.org/10.12942/lrr-2014-2} {\bibfield  {journal} {\bibinfo
  {journal} {Living Rev. Rel.}\ }\textbf {\bibinfo {volume} {17}},\ \bibinfo
  {pages} {2} (\bibinfo {year} {2014})},\ \Eprint
  {https://arxiv.org/abs/1310.1528} {arXiv:1310.1528 [gr-qc]} \BibitemShut
  {NoStop}%
\bibitem [{\citenamefont {Buonanno}\ and\ \citenamefont
  {Sathyaprakash}(2014)}]{Buonanno:2014aza}%
  \BibitemOpen
  \bibfield  {author} {\bibinfo {author} {\bibfnamefont {A.}~\bibnamefont
  {Buonanno}}\ and\ \bibinfo {author} {\bibfnamefont {B.~S.}\ \bibnamefont
  {Sathyaprakash}},\ }\bibinfo {title} {{Sources of Gravitational Waves: Theory
  and Observations}}\ (\bibinfo {year} {2014})\ \Eprint
  {https://arxiv.org/abs/1410.7832} {arXiv:1410.7832 [gr-qc]} \BibitemShut
  {NoStop}%
\bibitem [{\citenamefont {Porto}(2016)}]{Porto:2016pyg}%
  \BibitemOpen
  \bibfield  {author} {\bibinfo {author} {\bibfnamefont {R.~A.}\ \bibnamefont
  {Porto}},\ }\bibfield  {title} {\bibinfo {title} {{The effective field
  theorist\textquoteright{}s approach to gravitational dynamics}},\ }\href
  {https://doi.org/10.1016/j.physrep.2016.04.003} {\bibfield  {journal}
  {\bibinfo  {journal} {Phys. Rept.}\ }\textbf {\bibinfo {volume} {633}},\
  \bibinfo {pages} {1} (\bibinfo {year} {2016})},\ \Eprint
  {https://arxiv.org/abs/1601.04914} {arXiv:1601.04914 [hep-th]} \BibitemShut
  {NoStop}%
\bibitem [{\citenamefont {Blanchet}(1998)}]{Blanchet:1998in}%
  \BibitemOpen
  \bibfield  {author} {\bibinfo {author} {\bibfnamefont {L.}~\bibnamefont
  {Blanchet}},\ }\bibfield  {title} {\bibinfo {title} {{On the multipole
  expansion of the gravitational field}},\ }\href
  {https://doi.org/10.1088/0264-9381/15/7/013} {\bibfield  {journal} {\bibinfo
  {journal} {Class. Quant. Grav.}\ }\textbf {\bibinfo {volume} {15}},\ \bibinfo
  {pages} {1971} (\bibinfo {year} {1998})},\ \Eprint
  {https://arxiv.org/abs/gr-qc/9801101} {arXiv:gr-qc/9801101} \BibitemShut
  {NoStop}%
\bibitem [{\citenamefont {Blanchet}\ \emph
  {et~al.}(2023{\natexlab{a}})\citenamefont {Blanchet}, \citenamefont {Faye},
  \citenamefont {Henry}, \citenamefont {Larrouturou},\ and\ \citenamefont
  {Trestini}}]{Blanchet:2023sbv}%
  \BibitemOpen
  \bibfield  {author} {\bibinfo {author} {\bibfnamefont {L.}~\bibnamefont
  {Blanchet}}, \bibinfo {author} {\bibfnamefont {G.}~\bibnamefont {Faye}},
  \bibinfo {author} {\bibfnamefont {Q.}~\bibnamefont {Henry}}, \bibinfo
  {author} {\bibfnamefont {F.}~\bibnamefont {Larrouturou}},\ and\ \bibinfo
  {author} {\bibfnamefont {D.}~\bibnamefont {Trestini}},\ }\href@noop {}
  {\bibinfo {title} {{Gravitational Wave Flux and Quadrupole Modes from
  Quasi-Circular Non-Spinning Compact Binaries to the Fourth Post-Newtonian
  Order}}} (\bibinfo {year} {2023}{\natexlab{a}}),\ \Eprint
  {https://arxiv.org/abs/2304.11186} {arXiv:2304.11186 [gr-qc]} \BibitemShut
  {NoStop}%
\bibitem [{\citenamefont {Blanchet}\ \emph
  {et~al.}(2023{\natexlab{b}})\citenamefont {Blanchet}, \citenamefont {Faye},
  \citenamefont {Henry}, \citenamefont {Larrouturou},\ and\ \citenamefont
  {Trestini}}]{Blanchet:2023bwj}%
  \BibitemOpen
  \bibfield  {author} {\bibinfo {author} {\bibfnamefont {L.}~\bibnamefont
  {Blanchet}}, \bibinfo {author} {\bibfnamefont {G.}~\bibnamefont {Faye}},
  \bibinfo {author} {\bibfnamefont {Q.}~\bibnamefont {Henry}}, \bibinfo
  {author} {\bibfnamefont {F.}~\bibnamefont {Larrouturou}},\ and\ \bibinfo
  {author} {\bibfnamefont {D.}~\bibnamefont {Trestini}},\ }\href@noop {}
  {\bibinfo {title} {{Gravitational-Wave Phasing of Compact Binary Systems to
  the Fourth-and-a-Half post-Newtonian Order}}} (\bibinfo {year}
  {2023}{\natexlab{b}}),\ \Eprint {https://arxiv.org/abs/2304.11185}
  {arXiv:2304.11185 [gr-qc]} \BibitemShut {NoStop}%
\bibitem [{\citenamefont {Leibovich}\ \emph {et~al.}(2020)\citenamefont
  {Leibovich}, \citenamefont {Maia}, \citenamefont {Rothstein},\ and\
  \citenamefont {Yang}}]{Leibovich:2019cxo}%
  \BibitemOpen
  \bibfield  {author} {\bibinfo {author} {\bibfnamefont {A.~K.}\ \bibnamefont
  {Leibovich}}, \bibinfo {author} {\bibfnamefont {N.~T.}\ \bibnamefont {Maia}},
  \bibinfo {author} {\bibfnamefont {I.~Z.}\ \bibnamefont {Rothstein}},\ and\
  \bibinfo {author} {\bibfnamefont {Z.}~\bibnamefont {Yang}},\ }\bibfield
  {title} {\bibinfo {title} {{Second post-Newtonian order radiative dynamics of
  inspiralling compact binaries in the Effective Field Theory approach}},\
  }\href {https://doi.org/10.1103/PhysRevD.101.084058} {\bibfield  {journal}
  {\bibinfo  {journal} {Phys. Rev. D}\ }\textbf {\bibinfo {volume} {101}},\
  \bibinfo {pages} {084058} (\bibinfo {year} {2020})},\ \Eprint
  {https://arxiv.org/abs/1912.12546} {arXiv:1912.12546 [gr-qc]} \BibitemShut
  {NoStop}%
\bibitem [{\citenamefont {Jaranowski}\ and\ \citenamefont
  {Sch\"afer}(2013)}]{Jaranowski:2013lca}%
  \BibitemOpen
  \bibfield  {author} {\bibinfo {author} {\bibfnamefont {P.}~\bibnamefont
  {Jaranowski}}\ and\ \bibinfo {author} {\bibfnamefont {G.}~\bibnamefont
  {Sch\"afer}},\ }\bibfield  {title} {\bibinfo {title} {{Dimensional
  regularization of local singularities in the 4th post-Newtonian
  two-point-mass Hamiltonian}},\ }\href
  {https://doi.org/10.1103/PhysRevD.87.081503} {\bibfield  {journal} {\bibinfo
  {journal} {Phys. Rev. D}\ }\textbf {\bibinfo {volume} {87}},\ \bibinfo
  {pages} {081503} (\bibinfo {year} {2013})},\ \Eprint
  {https://arxiv.org/abs/1303.3225} {arXiv:1303.3225 [gr-qc]} \BibitemShut
  {NoStop}%
\bibitem [{\citenamefont {Jaranowski}\ and\ \citenamefont
  {Sch\"afer}(2015)}]{Jaranowski:2015lha}%
  \BibitemOpen
  \bibfield  {author} {\bibinfo {author} {\bibfnamefont {P.}~\bibnamefont
  {Jaranowski}}\ and\ \bibinfo {author} {\bibfnamefont {G.}~\bibnamefont
  {Sch\"afer}},\ }\bibfield  {title} {\bibinfo {title} {{Derivation of
  local-in-time fourth post-Newtonian ADM Hamiltonian for spinless compact
  binaries}},\ }\href {https://doi.org/10.1103/PhysRevD.92.124043} {\bibfield
  {journal} {\bibinfo  {journal} {Phys. Rev. D}\ }\textbf {\bibinfo {volume}
  {92}},\ \bibinfo {pages} {124043} (\bibinfo {year} {2015})},\ \Eprint
  {https://arxiv.org/abs/1508.01016} {arXiv:1508.01016 [gr-qc]} \BibitemShut
  {NoStop}%
\bibitem [{\citenamefont {Damour}\ \emph {et~al.}(2014)\citenamefont {Damour},
  \citenamefont {Jaranowski},\ and\ \citenamefont
  {Sch\"afer}}]{Damour:2014jta}%
  \BibitemOpen
  \bibfield  {author} {\bibinfo {author} {\bibfnamefont {T.}~\bibnamefont
  {Damour}}, \bibinfo {author} {\bibfnamefont {P.}~\bibnamefont {Jaranowski}},\
  and\ \bibinfo {author} {\bibfnamefont {G.}~\bibnamefont {Sch\"afer}},\
  }\bibfield  {title} {\bibinfo {title} {{Nonlocal-in-time action for the
  fourth post-Newtonian conservative dynamics of two-body systems}},\ }\href
  {https://doi.org/10.1103/PhysRevD.89.064058} {\bibfield  {journal} {\bibinfo
  {journal} {Phys. Rev. D}\ }\textbf {\bibinfo {volume} {89}},\ \bibinfo
  {pages} {064058} (\bibinfo {year} {2014})},\ \Eprint
  {https://arxiv.org/abs/1401.4548} {arXiv:1401.4548 [gr-qc]} \BibitemShut
  {NoStop}%
\bibitem [{\citenamefont {Bernard}\ \emph {et~al.}(2016)\citenamefont
  {Bernard}, \citenamefont {Blanchet}, \citenamefont {Boh\'e}, \citenamefont
  {Faye},\ and\ \citenamefont {Marsat}}]{Bernard:2015njp}%
  \BibitemOpen
  \bibfield  {author} {\bibinfo {author} {\bibfnamefont {L.}~\bibnamefont
  {Bernard}}, \bibinfo {author} {\bibfnamefont {L.}~\bibnamefont {Blanchet}},
  \bibinfo {author} {\bibfnamefont {A.}~\bibnamefont {Boh\'e}}, \bibinfo
  {author} {\bibfnamefont {G.}~\bibnamefont {Faye}},\ and\ \bibinfo {author}
  {\bibfnamefont {S.}~\bibnamefont {Marsat}},\ }\bibfield  {title} {\bibinfo
  {title} {{Fokker action of nonspinning compact binaries at the fourth
  post-Newtonian approximation}},\ }\href
  {https://doi.org/10.1103/PhysRevD.93.084037} {\bibfield  {journal} {\bibinfo
  {journal} {Phys. Rev. D}\ }\textbf {\bibinfo {volume} {93}},\ \bibinfo
  {pages} {084037} (\bibinfo {year} {2016})},\ \Eprint
  {https://arxiv.org/abs/1512.02876} {arXiv:1512.02876 [gr-qc]} \BibitemShut
  {NoStop}%
\bibitem [{\citenamefont {Bernard}\ \emph
  {et~al.}(2017{\natexlab{a}})\citenamefont {Bernard}, \citenamefont
  {Blanchet}, \citenamefont {Boh\'e}, \citenamefont {Faye},\ and\ \citenamefont
  {Marsat}}]{Bernard:2016wrg}%
  \BibitemOpen
  \bibfield  {author} {\bibinfo {author} {\bibfnamefont {L.}~\bibnamefont
  {Bernard}}, \bibinfo {author} {\bibfnamefont {L.}~\bibnamefont {Blanchet}},
  \bibinfo {author} {\bibfnamefont {A.}~\bibnamefont {Boh\'e}}, \bibinfo
  {author} {\bibfnamefont {G.}~\bibnamefont {Faye}},\ and\ \bibinfo {author}
  {\bibfnamefont {S.}~\bibnamefont {Marsat}},\ }\bibfield  {title} {\bibinfo
  {title} {{Energy and periastron advance of compact binaries on circular
  orbits at the fourth post-Newtonian order}},\ }\href
  {https://doi.org/10.1103/PhysRevD.95.044026} {\bibfield  {journal} {\bibinfo
  {journal} {Phys. Rev. D}\ }\textbf {\bibinfo {volume} {95}},\ \bibinfo
  {pages} {044026} (\bibinfo {year} {2017}{\natexlab{a}})},\ \Eprint
  {https://arxiv.org/abs/1610.07934} {arXiv:1610.07934 [gr-qc]} \BibitemShut
  {NoStop}%
\bibitem [{\citenamefont {Marchand}\ \emph {et~al.}(2018)\citenamefont
  {Marchand}, \citenamefont {Bernard}, \citenamefont {Blanchet},\ and\
  \citenamefont {Faye}}]{Marchand:2017pir}%
  \BibitemOpen
  \bibfield  {author} {\bibinfo {author} {\bibfnamefont {T.}~\bibnamefont
  {Marchand}}, \bibinfo {author} {\bibfnamefont {L.}~\bibnamefont {Bernard}},
  \bibinfo {author} {\bibfnamefont {L.}~\bibnamefont {Blanchet}},\ and\
  \bibinfo {author} {\bibfnamefont {G.}~\bibnamefont {Faye}},\ }\bibfield
  {title} {\bibinfo {title} {{Ambiguity-Free Completion of the Equations of
  Motion of Compact Binary Systems at the Fourth Post-Newtonian Order}},\
  }\href {https://doi.org/10.1103/PhysRevD.97.044023} {\bibfield  {journal}
  {\bibinfo  {journal} {Phys. Rev. D}\ }\textbf {\bibinfo {volume} {97}},\
  \bibinfo {pages} {044023} (\bibinfo {year} {2018})},\ \Eprint
  {https://arxiv.org/abs/1707.09289} {arXiv:1707.09289 [gr-qc]} \BibitemShut
  {NoStop}%
\bibitem [{\citenamefont {Bernard}\ \emph
  {et~al.}(2017{\natexlab{b}})\citenamefont {Bernard}, \citenamefont
  {Blanchet}, \citenamefont {Boh\'e}, \citenamefont {Faye},\ and\ \citenamefont
  {Marsat}}]{Bernard:2017bvn}%
  \BibitemOpen
  \bibfield  {author} {\bibinfo {author} {\bibfnamefont {L.}~\bibnamefont
  {Bernard}}, \bibinfo {author} {\bibfnamefont {L.}~\bibnamefont {Blanchet}},
  \bibinfo {author} {\bibfnamefont {A.}~\bibnamefont {Boh\'e}}, \bibinfo
  {author} {\bibfnamefont {G.}~\bibnamefont {Faye}},\ and\ \bibinfo {author}
  {\bibfnamefont {S.}~\bibnamefont {Marsat}},\ }\bibfield  {title} {\bibinfo
  {title} {{Dimensional regularization of the IR divergences in the Fokker
  action of point-particle binaries at the fourth post-Newtonian order}},\
  }\href {https://doi.org/10.1103/PhysRevD.96.104043} {\bibfield  {journal}
  {\bibinfo  {journal} {Phys. Rev. D}\ }\textbf {\bibinfo {volume} {96}},\
  \bibinfo {pages} {104043} (\bibinfo {year} {2017}{\natexlab{b}})},\ \Eprint
  {https://arxiv.org/abs/1706.08480} {arXiv:1706.08480 [gr-qc]} \BibitemShut
  {NoStop}%
\bibitem [{\citenamefont {Bernard}\ \emph {et~al.}(2018)\citenamefont
  {Bernard}, \citenamefont {Blanchet}, \citenamefont {Faye},\ and\
  \citenamefont {Marchand}}]{Bernard:2017ktp}%
  \BibitemOpen
  \bibfield  {author} {\bibinfo {author} {\bibfnamefont {L.}~\bibnamefont
  {Bernard}}, \bibinfo {author} {\bibfnamefont {L.}~\bibnamefont {Blanchet}},
  \bibinfo {author} {\bibfnamefont {G.}~\bibnamefont {Faye}},\ and\ \bibinfo
  {author} {\bibfnamefont {T.}~\bibnamefont {Marchand}},\ }\bibfield  {title}
  {\bibinfo {title} {{Center-of-Mass Equations of Motion and Conserved
  Integrals of Compact Binary Systems at the Fourth Post-Newtonian Order}},\
  }\href {https://doi.org/10.1103/PhysRevD.97.044037} {\bibfield  {journal}
  {\bibinfo  {journal} {Phys. Rev. D}\ }\textbf {\bibinfo {volume} {97}},\
  \bibinfo {pages} {044037} (\bibinfo {year} {2018})},\ \Eprint
  {https://arxiv.org/abs/1711.00283} {arXiv:1711.00283 [gr-qc]} \BibitemShut
  {NoStop}%
\bibitem [{\citenamefont {Goldberger}\ and\ \citenamefont
  {Rothstein}(2006)}]{Goldberger:2004jt}%
  \BibitemOpen
  \bibfield  {author} {\bibinfo {author} {\bibfnamefont {W.~D.}\ \bibnamefont
  {Goldberger}}\ and\ \bibinfo {author} {\bibfnamefont {I.~Z.}\ \bibnamefont
  {Rothstein}},\ }\bibfield  {title} {\bibinfo {title} {{An Effective field
  theory of gravity for extended objects}},\ }\href
  {https://doi.org/10.1103/PhysRevD.73.104029} {\bibfield  {journal} {\bibinfo
  {journal} {Phys. Rev. D}\ }\textbf {\bibinfo {volume} {73}},\ \bibinfo
  {pages} {104029} (\bibinfo {year} {2006})},\ \Eprint
  {https://arxiv.org/abs/hep-th/0409156} {arXiv:hep-th/0409156} \BibitemShut
  {NoStop}%
\bibitem [{\citenamefont {Foffa}\ and\ \citenamefont
  {Sturani}(2013{\natexlab{a}})}]{Foffa:2011np}%
  \BibitemOpen
  \bibfield  {author} {\bibinfo {author} {\bibfnamefont {S.}~\bibnamefont
  {Foffa}}\ and\ \bibinfo {author} {\bibfnamefont {R.}~\bibnamefont
  {Sturani}},\ }\bibfield  {title} {\bibinfo {title} {{Tail terms in
  gravitational radiation reaction via effective field theory}},\ }\href
  {https://doi.org/10.1103/PhysRevD.87.044056} {\bibfield  {journal} {\bibinfo
  {journal} {Phys. Rev. D}\ }\textbf {\bibinfo {volume} {87}},\ \bibinfo
  {pages} {044056} (\bibinfo {year} {2013}{\natexlab{a}})},\ \Eprint
  {https://arxiv.org/abs/1111.5488} {arXiv:1111.5488 [gr-qc]} \BibitemShut
  {NoStop}%
\bibitem [{\citenamefont {Foffa}\ and\ \citenamefont
  {Sturani}(2013{\natexlab{b}})}]{Foffa:2012rn}%
  \BibitemOpen
  \bibfield  {author} {\bibinfo {author} {\bibfnamefont {S.}~\bibnamefont
  {Foffa}}\ and\ \bibinfo {author} {\bibfnamefont {R.}~\bibnamefont
  {Sturani}},\ }\bibfield  {title} {\bibinfo {title} {{Dynamics of the
  gravitational two-body problem at fourth post-Newtonian order and at
  quadratic order in the Newton constant}},\ }\href
  {https://doi.org/10.1103/PhysRevD.87.064011} {\bibfield  {journal} {\bibinfo
  {journal} {Phys. Rev. D}\ }\textbf {\bibinfo {volume} {87}},\ \bibinfo
  {pages} {064011} (\bibinfo {year} {2013}{\natexlab{b}})},\ \Eprint
  {https://arxiv.org/abs/1206.7087} {arXiv:1206.7087 [gr-qc]} \BibitemShut
  {NoStop}%
\bibitem [{\citenamefont {Galley}\ \emph {et~al.}(2016)\citenamefont {Galley},
  \citenamefont {Leibovich}, \citenamefont {Porto},\ and\ \citenamefont
  {Ross}}]{Galley:2015kus}%
  \BibitemOpen
  \bibfield  {author} {\bibinfo {author} {\bibfnamefont {C.~R.}\ \bibnamefont
  {Galley}}, \bibinfo {author} {\bibfnamefont {A.~K.}\ \bibnamefont
  {Leibovich}}, \bibinfo {author} {\bibfnamefont {R.~A.}\ \bibnamefont
  {Porto}},\ and\ \bibinfo {author} {\bibfnamefont {A.}~\bibnamefont {Ross}},\
  }\bibfield  {title} {\bibinfo {title} {{Tail effect in gravitational
  radiation reaction: Time nonlocality and renormalization group evolution}},\
  }\href {https://doi.org/10.1103/PhysRevD.93.124010} {\bibfield  {journal}
  {\bibinfo  {journal} {Phys. Rev. D}\ }\textbf {\bibinfo {volume} {93}},\
  \bibinfo {pages} {124010} (\bibinfo {year} {2016})},\ \Eprint
  {https://arxiv.org/abs/1511.07379} {arXiv:1511.07379 [gr-qc]} \BibitemShut
  {NoStop}%
\bibitem [{\citenamefont {Foffa}\ \emph {et~al.}(2017)\citenamefont {Foffa},
  \citenamefont {Mastrolia}, \citenamefont {Sturani},\ and\ \citenamefont
  {Sturm}}]{Foffa:2016rgu}%
  \BibitemOpen
  \bibfield  {author} {\bibinfo {author} {\bibfnamefont {S.}~\bibnamefont
  {Foffa}}, \bibinfo {author} {\bibfnamefont {P.}~\bibnamefont {Mastrolia}},
  \bibinfo {author} {\bibfnamefont {R.}~\bibnamefont {Sturani}},\ and\ \bibinfo
  {author} {\bibfnamefont {C.}~\bibnamefont {Sturm}},\ }\bibfield  {title}
  {\bibinfo {title} {{Effective field theory approach to the gravitational
  two-body dynamics, at fourth post-Newtonian order and quintic in the Newton
  constant}},\ }\href {https://doi.org/10.1103/PhysRevD.95.104009} {\bibfield
  {journal} {\bibinfo  {journal} {Phys. Rev. D}\ }\textbf {\bibinfo {volume}
  {95}},\ \bibinfo {pages} {104009} (\bibinfo {year} {2017})},\ \Eprint
  {https://arxiv.org/abs/1612.00482} {arXiv:1612.00482 [gr-qc]} \BibitemShut
  {NoStop}%
\bibitem [{\citenamefont {Foffa}\ and\ \citenamefont
  {Sturani}(2019)}]{Foffa:2019rdf}%
  \BibitemOpen
  \bibfield  {author} {\bibinfo {author} {\bibfnamefont {S.}~\bibnamefont
  {Foffa}}\ and\ \bibinfo {author} {\bibfnamefont {R.}~\bibnamefont
  {Sturani}},\ }\bibfield  {title} {\bibinfo {title} {{Conservative dynamics of
  binary systems to fourth Post-Newtonian order in the EFT approach I:
  Regularized Lagrangian}},\ }\href
  {https://doi.org/10.1103/PhysRevD.100.024047} {\bibfield  {journal} {\bibinfo
   {journal} {Phys. Rev. D}\ }\textbf {\bibinfo {volume} {100}},\ \bibinfo
  {pages} {024047} (\bibinfo {year} {2019})},\ \Eprint
  {https://arxiv.org/abs/1903.05113} {arXiv:1903.05113 [gr-qc]} \BibitemShut
  {NoStop}%
\bibitem [{\citenamefont {Foffa}\ \emph
  {et~al.}(2019{\natexlab{a}})\citenamefont {Foffa}, \citenamefont {Porto},
  \citenamefont {Rothstein},\ and\ \citenamefont {Sturani}}]{Foffa:2019yfl}%
  \BibitemOpen
  \bibfield  {author} {\bibinfo {author} {\bibfnamefont {S.}~\bibnamefont
  {Foffa}}, \bibinfo {author} {\bibfnamefont {R.~A.}\ \bibnamefont {Porto}},
  \bibinfo {author} {\bibfnamefont {I.}~\bibnamefont {Rothstein}},\ and\
  \bibinfo {author} {\bibfnamefont {R.}~\bibnamefont {Sturani}},\ }\bibfield
  {title} {\bibinfo {title} {{Conservative dynamics of binary systems to fourth
  Post-Newtonian order in the EFT approach II: Renormalized Lagrangian}},\
  }\href {https://doi.org/10.1103/PhysRevD.100.024048} {\bibfield  {journal}
  {\bibinfo  {journal} {Phys. Rev. D}\ }\textbf {\bibinfo {volume} {100}},\
  \bibinfo {pages} {024048} (\bibinfo {year} {2019}{\natexlab{a}})},\ \Eprint
  {https://arxiv.org/abs/1903.05118} {arXiv:1903.05118 [gr-qc]} \BibitemShut
  {NoStop}%
\bibitem [{\citenamefont {Bl\"umlein}\ \emph {et~al.}(2020)\citenamefont
  {Bl\"umlein}, \citenamefont {Maier}, \citenamefont {Marquard},\ and\
  \citenamefont {Sch\"afer}}]{Blumlein:2020pog}%
  \BibitemOpen
  \bibfield  {author} {\bibinfo {author} {\bibfnamefont {J.}~\bibnamefont
  {Bl\"umlein}}, \bibinfo {author} {\bibfnamefont {A.}~\bibnamefont {Maier}},
  \bibinfo {author} {\bibfnamefont {P.}~\bibnamefont {Marquard}},\ and\
  \bibinfo {author} {\bibfnamefont {G.}~\bibnamefont {Sch\"afer}},\ }\bibfield
  {title} {\bibinfo {title} {{Fourth post-Newtonian Hamiltonian dynamics of
  two-body systems from an effective field theory approach}},\ }\href
  {https://doi.org/10.1016/j.nuclphysb.2020.115041} {\bibfield  {journal}
  {\bibinfo  {journal} {Nucl. Phys. B}\ }\textbf {\bibinfo {volume} {955}},\
  \bibinfo {pages} {115041} (\bibinfo {year} {2020})},\ \Eprint
  {https://arxiv.org/abs/2003.01692} {arXiv:2003.01692 [gr-qc]} \BibitemShut
  {NoStop}%
\bibitem [{\citenamefont {Foffa}\ \emph
  {et~al.}(2019{\natexlab{b}})\citenamefont {Foffa}, \citenamefont {Mastrolia},
  \citenamefont {Sturani}, \citenamefont {Sturm},\ and\ \citenamefont
  {Torres~Bobadilla}}]{Foffa:2019hrb}%
  \BibitemOpen
  \bibfield  {author} {\bibinfo {author} {\bibfnamefont {S.}~\bibnamefont
  {Foffa}}, \bibinfo {author} {\bibfnamefont {P.}~\bibnamefont {Mastrolia}},
  \bibinfo {author} {\bibfnamefont {R.}~\bibnamefont {Sturani}}, \bibinfo
  {author} {\bibfnamefont {C.}~\bibnamefont {Sturm}},\ and\ \bibinfo {author}
  {\bibfnamefont {W.~J.}\ \bibnamefont {Torres~Bobadilla}},\ }\bibfield
  {title} {\bibinfo {title} {{Static two-body potential at fifth post-Newtonian
  order}},\ }\href {https://doi.org/10.1103/PhysRevLett.122.241605} {\bibfield
  {journal} {\bibinfo  {journal} {Phys. Rev. Lett.}\ }\textbf {\bibinfo
  {volume} {122}},\ \bibinfo {pages} {241605} (\bibinfo {year}
  {2019}{\natexlab{b}})},\ \Eprint {https://arxiv.org/abs/1902.10571}
  {arXiv:1902.10571 [gr-qc]} \BibitemShut {NoStop}%
\bibitem [{\citenamefont {Bini}\ \emph {et~al.}(2019)\citenamefont {Bini},
  \citenamefont {Damour},\ and\ \citenamefont {Geralico}}]{Bini:2019nra}%
  \BibitemOpen
  \bibfield  {author} {\bibinfo {author} {\bibfnamefont {D.}~\bibnamefont
  {Bini}}, \bibinfo {author} {\bibfnamefont {T.}~\bibnamefont {Damour}},\ and\
  \bibinfo {author} {\bibfnamefont {A.}~\bibnamefont {Geralico}},\ }\bibfield
  {title} {\bibinfo {title} {{Novel approach to binary dynamics: application to
  the fifth post-Newtonian level}},\ }\href
  {https://doi.org/10.1103/PhysRevLett.123.231104} {\bibfield  {journal}
  {\bibinfo  {journal} {Phys. Rev. Lett.}\ }\textbf {\bibinfo {volume} {123}},\
  \bibinfo {pages} {231104} (\bibinfo {year} {2019})},\ \Eprint
  {https://arxiv.org/abs/1909.02375} {arXiv:1909.02375 [gr-qc]} \BibitemShut
  {NoStop}%
\bibitem [{\citenamefont {Bini}\ \emph
  {et~al.}(2020{\natexlab{a}})\citenamefont {Bini}, \citenamefont {Damour},\
  and\ \citenamefont {Geralico}}]{Bini:2020wpo}%
  \BibitemOpen
  \bibfield  {author} {\bibinfo {author} {\bibfnamefont {D.}~\bibnamefont
  {Bini}}, \bibinfo {author} {\bibfnamefont {T.}~\bibnamefont {Damour}},\ and\
  \bibinfo {author} {\bibfnamefont {A.}~\bibnamefont {Geralico}},\ }\bibfield
  {title} {\bibinfo {title} {{Binary dynamics at the fifth and fifth-and-a-half
  post-Newtonian orders}},\ }\href
  {https://doi.org/10.1103/PhysRevD.102.024062} {\bibfield  {journal} {\bibinfo
   {journal} {Phys. Rev. D}\ }\textbf {\bibinfo {volume} {102}},\ \bibinfo
  {pages} {024062} (\bibinfo {year} {2020}{\natexlab{a}})},\ \Eprint
  {https://arxiv.org/abs/2003.11891} {arXiv:2003.11891 [gr-qc]} \BibitemShut
  {NoStop}%
\bibitem [{\citenamefont {Bl\"umlein}\ \emph {et~al.}(2021)\citenamefont
  {Bl\"umlein}, \citenamefont {Maier}, \citenamefont {Marquard},\ and\
  \citenamefont {Sch\"afer}}]{Blumlein:2020pyo}%
  \BibitemOpen
  \bibfield  {author} {\bibinfo {author} {\bibfnamefont {J.}~\bibnamefont
  {Bl\"umlein}}, \bibinfo {author} {\bibfnamefont {A.}~\bibnamefont {Maier}},
  \bibinfo {author} {\bibfnamefont {P.}~\bibnamefont {Marquard}},\ and\
  \bibinfo {author} {\bibfnamefont {G.}~\bibnamefont {Sch\"afer}},\ }\bibfield
  {title} {\bibinfo {title} {{The fifth-order post-Newtonian Hamiltonian
  dynamics of two-body systems from an effective field theory approach:
  potential contributions}},\ }\href
  {https://doi.org/10.1016/j.nuclphysb.2021.115352} {\bibfield  {journal}
  {\bibinfo  {journal} {Nucl. Phys. B}\ }\textbf {\bibinfo {volume} {965}},\
  \bibinfo {pages} {115352} (\bibinfo {year} {2021})},\ \Eprint
  {https://arxiv.org/abs/2010.13672} {arXiv:2010.13672 [gr-qc]} \BibitemShut
  {NoStop}%
\bibitem [{\citenamefont {Bini}\ \emph
  {et~al.}(2020{\natexlab{b}})\citenamefont {Bini}, \citenamefont {Damour},\
  and\ \citenamefont {Geralico}}]{Bini:2020nsb}%
  \BibitemOpen
  \bibfield  {author} {\bibinfo {author} {\bibfnamefont {D.}~\bibnamefont
  {Bini}}, \bibinfo {author} {\bibfnamefont {T.}~\bibnamefont {Damour}},\ and\
  \bibinfo {author} {\bibfnamefont {A.}~\bibnamefont {Geralico}},\ }\bibfield
  {title} {\bibinfo {title} {{Sixth post-Newtonian local-in-time dynamics of
  binary systems}},\ }\href {https://doi.org/10.1103/PhysRevD.102.024061}
  {\bibfield  {journal} {\bibinfo  {journal} {Phys. Rev. D}\ }\textbf {\bibinfo
  {volume} {102}},\ \bibinfo {pages} {024061} (\bibinfo {year}
  {2020}{\natexlab{b}})},\ \Eprint {https://arxiv.org/abs/2004.05407}
  {arXiv:2004.05407 [gr-qc]} \BibitemShut {NoStop}%
\bibitem [{\citenamefont {Bini}\ \emph {et~al.}(2021)\citenamefont {Bini},
  \citenamefont {Damour}, \citenamefont {Geralico}, \citenamefont {Laporta},\
  and\ \citenamefont {Mastrolia}}]{Bini:2020rzn}%
  \BibitemOpen
  \bibfield  {author} {\bibinfo {author} {\bibfnamefont {D.}~\bibnamefont
  {Bini}}, \bibinfo {author} {\bibfnamefont {T.}~\bibnamefont {Damour}},
  \bibinfo {author} {\bibfnamefont {A.}~\bibnamefont {Geralico}}, \bibinfo
  {author} {\bibfnamefont {S.}~\bibnamefont {Laporta}},\ and\ \bibinfo {author}
  {\bibfnamefont {P.}~\bibnamefont {Mastrolia}},\ }\bibfield  {title} {\bibinfo
  {title} {{Gravitational scattering at the seventh order in $G$: nonlocal
  contribution at the sixth post-Newtonian accuracy}},\ }\href
  {https://doi.org/10.1103/PhysRevD.103.044038} {\bibfield  {journal} {\bibinfo
   {journal} {Phys. Rev. D}\ }\textbf {\bibinfo {volume} {103}},\ \bibinfo
  {pages} {044038} (\bibinfo {year} {2021})},\ \Eprint
  {https://arxiv.org/abs/2012.12918} {arXiv:2012.12918 [gr-qc]} \BibitemShut
  {NoStop}%
\bibitem [{\citenamefont {Almeida}\ \emph {et~al.}(2021)\citenamefont
  {Almeida}, \citenamefont {Foffa},\ and\ \citenamefont
  {Sturani}}]{Almeida:2021xwn}%
  \BibitemOpen
  \bibfield  {author} {\bibinfo {author} {\bibfnamefont {G.~L.}\ \bibnamefont
  {Almeida}}, \bibinfo {author} {\bibfnamefont {S.}~\bibnamefont {Foffa}},\
  and\ \bibinfo {author} {\bibfnamefont {R.}~\bibnamefont {Sturani}},\
  }\bibfield  {title} {\bibinfo {title} {{Tail contributions to gravitational
  conservative dynamics}},\ }\href
  {https://doi.org/10.1103/PhysRevD.104.124075} {\bibfield  {journal} {\bibinfo
   {journal} {Phys. Rev. D}\ }\textbf {\bibinfo {volume} {104}},\ \bibinfo
  {pages} {124075} (\bibinfo {year} {2021})},\ \Eprint
  {https://arxiv.org/abs/2110.14146} {arXiv:2110.14146 [gr-qc]} \BibitemShut
  {NoStop}%
\bibitem [{\citenamefont {Foffa}\ and\ \citenamefont
  {Sturani}(2020)}]{Foffa:2019eeb}%
  \BibitemOpen
  \bibfield  {author} {\bibinfo {author} {\bibfnamefont {S.}~\bibnamefont
  {Foffa}}\ and\ \bibinfo {author} {\bibfnamefont {R.}~\bibnamefont
  {Sturani}},\ }\bibfield  {title} {\bibinfo {title} {{Hereditary terms at
  next-to-leading order in two-body gravitational dynamics}},\ }\href
  {https://doi.org/10.1103/PhysRevD.101.064033} {\bibfield  {journal} {\bibinfo
   {journal} {Phys. Rev. D}\ }\textbf {\bibinfo {volume} {101}},\ \bibinfo
  {pages} {064033} (\bibinfo {year} {2020})},\ \bibinfo {note} {[Erratum:
  Phys.Rev.D 103, 089901 (2021)]},\ \Eprint {https://arxiv.org/abs/1907.02869}
  {arXiv:1907.02869 [gr-qc]} \BibitemShut {NoStop}%
\bibitem [{\citenamefont {Bl\"umlein}\ \emph {et~al.}(2022)\citenamefont
  {Bl\"umlein}, \citenamefont {Maier}, \citenamefont {Marquard},\ and\
  \citenamefont {Sch\"afer}}]{Blumlein:2021txe}%
  \BibitemOpen
  \bibfield  {author} {\bibinfo {author} {\bibfnamefont {J.}~\bibnamefont
  {Bl\"umlein}}, \bibinfo {author} {\bibfnamefont {A.}~\bibnamefont {Maier}},
  \bibinfo {author} {\bibfnamefont {P.}~\bibnamefont {Marquard}},\ and\
  \bibinfo {author} {\bibfnamefont {G.}~\bibnamefont {Sch\"afer}},\ }\bibfield
  {title} {\bibinfo {title} {{The fifth-order post-Newtonian Hamiltonian
  dynamics of two-body systems from an effective field theory approach}},\
  }\href {https://doi.org/10.1016/j.nuclphysb.2022.115900} {\bibfield
  {journal} {\bibinfo  {journal} {Nucl. Phys. B}\ }\textbf {\bibinfo {volume}
  {983}},\ \bibinfo {pages} {115900} (\bibinfo {year} {2022})},\ \bibinfo
  {note} {[Erratum: Nucl.Phys.B 985, 115991 (2022)]},\ \Eprint
  {https://arxiv.org/abs/2110.13822} {arXiv:2110.13822 [gr-qc]} \BibitemShut
  {NoStop}%
\bibitem [{\citenamefont {Almeida}\ \emph {et~al.}(2023)\citenamefont
  {Almeida}, \citenamefont {M\"uller}, \citenamefont {Foffa},\ and\
  \citenamefont {Sturani}}]{Gabriel}%
  \BibitemOpen
  \bibfield  {author} {\bibinfo {author} {\bibfnamefont {G.~L.}\ \bibnamefont
  {Almeida}}, \bibinfo {author} {\bibfnamefont {A.}~\bibnamefont {M\"uller}},
  \bibinfo {author} {\bibfnamefont {S.}~\bibnamefont {Foffa}},\ and\ \bibinfo
  {author} {\bibfnamefont {R.}~\bibnamefont {Sturani}},\ }\href@noop {}
  {\bibinfo {title} {{Conservative binary dynamics from gravitational tail
  emission processes}}} (\bibinfo {year} {2023}),\ \Eprint
  {https://arxiv.org/abs/2307.05327} {arXiv:2307.05327 [gr-qc]} \BibitemShut
  {NoStop}%
\bibitem [{\citenamefont {Blanchet}\ \emph {et~al.}(2020)\citenamefont
  {Blanchet}, \citenamefont {Foffa}, \citenamefont {Larrouturou},\ and\
  \citenamefont {Sturani}}]{Blanchet:2019rjs}%
  \BibitemOpen
  \bibfield  {author} {\bibinfo {author} {\bibfnamefont {L.}~\bibnamefont
  {Blanchet}}, \bibinfo {author} {\bibfnamefont {S.}~\bibnamefont {Foffa}},
  \bibinfo {author} {\bibfnamefont {F.}~\bibnamefont {Larrouturou}},\ and\
  \bibinfo {author} {\bibfnamefont {R.}~\bibnamefont {Sturani}},\ }\bibfield
  {title} {\bibinfo {title} {{Logarithmic tail contributions to the energy
  function of circular compact binaries}},\ }\href
  {https://doi.org/10.1103/PhysRevD.101.084045} {\bibfield  {journal} {\bibinfo
   {journal} {Phys. Rev. D}\ }\textbf {\bibinfo {volume} {101}},\ \bibinfo
  {pages} {084045} (\bibinfo {year} {2020})},\ \Eprint
  {https://arxiv.org/abs/1912.12359} {arXiv:1912.12359 [gr-qc]} \BibitemShut
  {NoStop}%
\bibitem [{\citenamefont {Henry}\ \emph {et~al.}(2021)\citenamefont {Henry},
  \citenamefont {Faye},\ and\ \citenamefont {Blanchet}}]{Henry:2021cek}%
  \BibitemOpen
  \bibfield  {author} {\bibinfo {author} {\bibfnamefont {Q.}~\bibnamefont
  {Henry}}, \bibinfo {author} {\bibfnamefont {G.}~\bibnamefont {Faye}},\ and\
  \bibinfo {author} {\bibfnamefont {L.}~\bibnamefont {Blanchet}},\ }\bibfield
  {title} {\bibinfo {title} {{The current-type quadrupole moment and
  gravitational-wave mode (\ensuremath{\ell}, m) = (2, 1) of compact binary
  systems at the third post-Newtonian order}},\ }\href
  {https://doi.org/10.1088/1361-6382/ac1850} {\bibfield  {journal} {\bibinfo
  {journal} {Class. Quant. Grav.}\ }\textbf {\bibinfo {volume} {38}},\ \bibinfo
  {pages} {185004} (\bibinfo {year} {2021})},\ \Eprint
  {https://arxiv.org/abs/2105.10876} {arXiv:2105.10876 [gr-qc]} \BibitemShut
  {NoStop}%
\bibitem [{\citenamefont {Marchand}\ \emph {et~al.}(2020)\citenamefont
  {Marchand}, \citenamefont {Henry}, \citenamefont {Larrouturou}, \citenamefont
  {Marsat}, \citenamefont {Faye},\ and\ \citenamefont
  {Blanchet}}]{Marchand:2020fpt}%
  \BibitemOpen
  \bibfield  {author} {\bibinfo {author} {\bibfnamefont {T.}~\bibnamefont
  {Marchand}}, \bibinfo {author} {\bibfnamefont {Q.}~\bibnamefont {Henry}},
  \bibinfo {author} {\bibfnamefont {F.}~\bibnamefont {Larrouturou}}, \bibinfo
  {author} {\bibfnamefont {S.}~\bibnamefont {Marsat}}, \bibinfo {author}
  {\bibfnamefont {G.}~\bibnamefont {Faye}},\ and\ \bibinfo {author}
  {\bibfnamefont {L.}~\bibnamefont {Blanchet}},\ }\bibfield  {title} {\bibinfo
  {title} {{The mass quadrupole moment of compact binary systems at the fourth
  post-Newtonian order}},\ }\href {https://doi.org/10.1088/1361-6382/ab9ce1}
  {\bibfield  {journal} {\bibinfo  {journal} {Class. Quant. Grav.}\ }\textbf
  {\bibinfo {volume} {37}},\ \bibinfo {pages} {215006} (\bibinfo {year}
  {2020})},\ \Eprint {https://arxiv.org/abs/2003.13672} {arXiv:2003.13672
  [gr-qc]} \BibitemShut {NoStop}%
\bibitem [{\citenamefont {Larrouturou}\ \emph
  {et~al.}(2022{\natexlab{a}})\citenamefont {Larrouturou}, \citenamefont
  {Henry}, \citenamefont {Blanchet},\ and\ \citenamefont
  {Faye}}]{Larrouturou:2021dma}%
  \BibitemOpen
  \bibfield  {author} {\bibinfo {author} {\bibfnamefont {F.}~\bibnamefont
  {Larrouturou}}, \bibinfo {author} {\bibfnamefont {Q.}~\bibnamefont {Henry}},
  \bibinfo {author} {\bibfnamefont {L.}~\bibnamefont {Blanchet}},\ and\
  \bibinfo {author} {\bibfnamefont {G.}~\bibnamefont {Faye}},\ }\bibfield
  {title} {\bibinfo {title} {{The quadrupole moment of compact binaries to the
  fourth post-Newtonian order: I. Non-locality in time and infra-red
  divergencies}},\ }\href {https://doi.org/10.1088/1361-6382/ac5762} {\bibfield
   {journal} {\bibinfo  {journal} {Class. Quant. Grav.}\ }\textbf {\bibinfo
  {volume} {39}},\ \bibinfo {pages} {115007} (\bibinfo {year}
  {2022}{\natexlab{a}})},\ \Eprint {https://arxiv.org/abs/2110.02240}
  {arXiv:2110.02240 [gr-qc]} \BibitemShut {NoStop}%
\bibitem [{\citenamefont {Thorne}(1980)}]{Thorne:1980ru}%
  \BibitemOpen
  \bibfield  {author} {\bibinfo {author} {\bibfnamefont {K.~S.}\ \bibnamefont
  {Thorne}},\ }\bibfield  {title} {\bibinfo {title} {{Multipole Expansions of
  Gravitational Radiation}},\ }\href
  {https://doi.org/10.1103/RevModPhys.52.299} {\bibfield  {journal} {\bibinfo
  {journal} {Rev. Mod. Phys.}\ }\textbf {\bibinfo {volume} {52}},\ \bibinfo
  {pages} {299} (\bibinfo {year} {1980})}\BibitemShut {NoStop}%
\bibitem [{\citenamefont {Blanchet}\ and\ \citenamefont
  {Faye}(2000)}]{Blanchet:2000nu}%
  \BibitemOpen
  \bibfield  {author} {\bibinfo {author} {\bibfnamefont {L.}~\bibnamefont
  {Blanchet}}\ and\ \bibinfo {author} {\bibfnamefont {G.}~\bibnamefont
  {Faye}},\ }\bibfield  {title} {\bibinfo {title} {{Hadamard regularization}},\
  }\href {https://doi.org/10.1063/1.1308506} {\bibfield  {journal} {\bibinfo
  {journal} {J. Math. Phys.}\ }\textbf {\bibinfo {volume} {41}},\ \bibinfo
  {pages} {7675} (\bibinfo {year} {2000})},\ \Eprint
  {https://arxiv.org/abs/gr-qc/0004008} {arXiv:gr-qc/0004008} \BibitemShut
  {NoStop}%
\bibitem [{\citenamefont {Blanchet}(1993)}]{Blanchet:1993ng}%
  \BibitemOpen
  \bibfield  {author} {\bibinfo {author} {\bibfnamefont {L.}~\bibnamefont
  {Blanchet}},\ }\bibfield  {title} {\bibinfo {title} {{Time asymmetric
  structure of gravitational radiation}},\ }\href
  {https://doi.org/10.1103/PhysRevD.47.4392} {\bibfield  {journal} {\bibinfo
  {journal} {Phys. Rev. D}\ }\textbf {\bibinfo {volume} {47}},\ \bibinfo
  {pages} {4392} (\bibinfo {year} {1993})}\BibitemShut {NoStop}%
\bibitem [{\citenamefont {Larrouturou}\ \emph
  {et~al.}(2022{\natexlab{b}})\citenamefont {Larrouturou}, \citenamefont
  {Blanchet}, \citenamefont {Henry},\ and\ \citenamefont
  {Faye}}]{Larrouturou:2021gqo}%
  \BibitemOpen
  \bibfield  {author} {\bibinfo {author} {\bibfnamefont {F.}~\bibnamefont
  {Larrouturou}}, \bibinfo {author} {\bibfnamefont {L.}~\bibnamefont
  {Blanchet}}, \bibinfo {author} {\bibfnamefont {Q.}~\bibnamefont {Henry}},\
  and\ \bibinfo {author} {\bibfnamefont {G.}~\bibnamefont {Faye}},\ }\bibfield
  {title} {\bibinfo {title} {{The quadrupole moment of compact binaries to the
  fourth post-Newtonian order: II. Dimensional regularization and
  renormalization}},\ }\href {https://doi.org/10.1088/1361-6382/ac5ba0}
  {\bibfield  {journal} {\bibinfo  {journal} {Class. Quant. Grav.}\ }\textbf
  {\bibinfo {volume} {39}},\ \bibinfo {pages} {115008} (\bibinfo {year}
  {2022}{\natexlab{b}})},\ \Eprint {https://arxiv.org/abs/2110.02243}
  {arXiv:2110.02243 [gr-qc]} \BibitemShut {NoStop}%
\bibitem [{\citenamefont {Mart\'in-Garc\'ia}\ \emph {et~al.}(2012)\citenamefont
  {Mart\'in-Garc\'ia}, \citenamefont {Garc\'ia-Parrado}, \citenamefont
  {Stecchina}, \citenamefont {Wardell}, \citenamefont {Pitrou}, \citenamefont
  {Brizuela}, \citenamefont {Yllanes}, \citenamefont {Faye}, \citenamefont
  {Stein}, \citenamefont {Portugal},\ and\ \citenamefont
  {B\"ackdahl}}]{xtensor}%
  \BibitemOpen
  \bibfield  {author} {\bibinfo {author} {\bibfnamefont {J.~M.}\ \bibnamefont
  {Mart\'in-Garc\'ia}}, \bibinfo {author} {\bibfnamefont {A.}~\bibnamefont
  {Garc\'ia-Parrado}}, \bibinfo {author} {\bibfnamefont {A.}~\bibnamefont
  {Stecchina}}, \bibinfo {author} {\bibfnamefont {B.}~\bibnamefont {Wardell}},
  \bibinfo {author} {\bibfnamefont {C.}~\bibnamefont {Pitrou}}, \bibinfo
  {author} {\bibfnamefont {D.}~\bibnamefont {Brizuela}}, \bibinfo {author}
  {\bibfnamefont {D.}~\bibnamefont {Yllanes}}, \bibinfo {author} {\bibfnamefont
  {G.}~\bibnamefont {Faye}}, \bibinfo {author} {\bibfnamefont {L.}~\bibnamefont
  {Stein}}, \bibinfo {author} {\bibfnamefont {R.}~\bibnamefont {Portugal}},\
  and\ \bibinfo {author} {\bibfnamefont {T.}~\bibnamefont {B\"ackdahl}},\
  }\href@noop {} {\bibinfo {title} {{xAct}: Efficient tensor computer algebra
  for {Mathematica}}} (\bibinfo {year} {GPL 2002--2012}),\ \bibinfo {note}
  {http://www.xact.es/}\BibitemShut {NoStop}%
\end{thebibliography}%

\end{document}